\documentclass[pre,aps,singlecolumn,12pt]{revtex4-1}
\usepackage[latin1]{inputenc}
\usepackage[english]{babel}
\usepackage{graphicx}
\usepackage{color}
\usepackage{amsmath}
\usepackage{amssymb}

\begin{document}

\title{\bf On non--dissipative and dissipative qubit manifolds}
\author{H. C. Pe\~nate--Rodr\'{\i}guez $^{1}$, P. Bargue\~no $^{2,*}$, G. Rojas--Lorenzo $^{1}$ and
S. Miret--Art\'es $^{3}$
}

\affiliation{$^{1}$
Instituto Superior de Tecnolog\'{\i}as y Ciencias Aplicadas, Ave. Salvador Allende y Luaces,
Quinta de los Molinos, Plaza, La Habana {\it 10600}, Cuba
\\
$^{2}$
Departamento de F\'{\i}sica de Materiales,
Universidad Complutense de Madrid, {\it 28040}, Madrid, Spain
\\
$^{3}$ Instituto de F\'{\i}sica Fundamental (CSIC), Serrano 123, {\it 28006}, Madrid, Spain
\\
($*$ p.bargueno@fis.ucm.es)
}
\begin{abstract}

The trajectories of a qubit dynamics over the two--sphere are
shown to be geodesics of certain Riemannian or physically--sound
Lorentzian manifolds, both in the non--dissipative and dissipative
formalisms, when using action--angle variables. Several aspects of
the geometry and topology of these manifolds (qubit manifolds)
have been studied for some special physical cases.
\end{abstract}

\maketitle

\section{Introduction}

Since its conception, one of the paradigms of quantum mechanics is
the two--level system. Its role in almost all the fields of
physics is difficult to overemphasize, going, for instance, from
high energy physics \cite{Rosner2001} and parity--violating chiral
molecules \cite{Harris1978,Quack1986} to macroscopic quantum
phenomena, as shown by Feynman in his elegant and pedagogical
dynamical theory of the Josephson effect \cite{Feynmanbook}.
During the past years, the interest in two--level systems has
increased considerably due to its applicability in quantum
computation under the name of {\it qubits}
\cite{Ladd2010,Nielsenbook}. Interestingly, not only for
physicists but also for mathematicians the qubit can be used to
explore and test sophisticated theories, in particular using
geometrical concepts. Specifically, due to the usual decomposition
of the scalar product between two states in the associated Hilbert
space, $\mathcal{H}$, in its real and imaginary parts, both
Riemannian and symplectic structures can be introduced in
$\mathcal{H}$, which turns out to be the basis for the
geometrization of quantum mechanics. Although the probabilistic
aspects of the theory, including the uncertainty principle and
related facts are due to the Riemannian structure, the whole
quantum dynamics can be formulated as a pure classical theory by
defining a symplectic structure over the projective Hilbert space,
$\mathcal{P}(\mathcal{H})$, taken as a K\"ahler manifold, which is
the quantum phase space where the dynamics takes place. These
interesting points and subsequent extensions were made by Kibble
\cite{Kibble1979} and other authors
\cite{Heslot1985,Anandan1990,Gibbons1992,Ashtekar1998,Brody2001},
respectively (for a very readable introduction to geometric
structures in quantum mechanics see, for
example,\cite{Chruszinskybook}).

For our purposes, let us briefly sketch this quantum--classical
equivalence for a qubit. If $\mathcal{H}$ is a two--dimensional
Hilbert space and $|\Psi\rangle \in \mathcal{H}$ is a normalized
qubit, then $|\Psi\rangle \in S^{3}$. By the celebrated first Hopf
fibration $S^{3}\to S^{2}$
\cite{Hopf1931,Urbantke1991,Mosseri2001,Urbantke2003} we can gauge
out the global phase and arrive at the Bloch sphere
representation. This map can be understood as a composition,
$\Pi=\Xi  \circ \Omega$, where $\Omega:S^{3}\subset
\mathbb{C}^{2}\to \mathbb{C}P^{1}$ links an element of
$\mathbb{C}^{2}$ to its equivalence class and
$\Xi:\mathbb{C}P^{1}\left(=\mathbb{C}\cup \infty \right)\to S^{2}$
is given by the stereographic projection. It can be shown that the
Hopf map can be written in terms of the Pauli matrices as
$\Pi\left(|\Psi\rangle \in S^{3}\right)= \left(\langle
\Psi|\hat\sigma_{x}|\Psi\rangle,\langle
\Psi|\hat\sigma_{y}|\Psi\rangle,\langle
\Psi|\hat\sigma_{z}|\Psi\rangle \right) \in S^{2}$, where $\langle
\Psi|\hat\sigma_{x}|\Psi\rangle^{2}+\langle
\Psi|\hat\sigma_{y}|\Psi\rangle^{2} +\langle
\Psi|\hat\sigma_{z}|\Psi\rangle^{2}=1$. Thus, from the first Hopf
map, quantum and classical mechanics may be embedded in the same
formulation. Specifically, for the qubit case, the Strocchi map
\cite{Strocchi1966} is exactly the Hopf map previously described.
After defining appropriate canonical, action--angle variables ($I,
\Phi$) on $S^{2}$, a classical Hamiltonian function can be
derived. In fact, one can prove that the Schr\"odinger dynamics on
$\mathcal{H}$ corresponds to a Hamiltonian dynamics defined by the
symplectic form $\Omega=d\Phi\wedge dI$ on $S^{2}$. Thus, $S^{2}$,
taken as a symplectic manifold, can be regarded as the quantum
phase space for a qubit.

In the two--dimensional case, the normalized qubit state can be
expanded as $|\Psi\rangle=a_{1}|1\rangle +a_{2}|2\rangle$, where
$a_{j}=|a_{j}|e^{i\phi_{j}}\in \mathbb{C}$. Let us define the pair
of action--angle variables as $I \equiv |a_{1}|^{2}-|a_{2}|^{2}$
and $\Phi\equiv \phi_{1}-\phi_{2}$. Then, a general Hamiltonian operator
$\hat H= \sum_{i}A^{i}\hat \sigma_{i}$, where $\hat\sigma_{i}$ are
the Pauli matrices and $A^{i}\in \mathbb{R}$, can be mapped into
the Hamiltonian function
\begin{eqnarray}
H_{0}&=&2\langle \Psi| \hat H|\Psi\rangle \nonumber \\
&=&\sqrt{1-I^{2}}\left(2A_{x}\cos\Phi +2A_{y}\sin\Phi\right) + 2
A_{z}I
\end{eqnarray}
where $H_0$ is a generalized Meyer--Miller-Stock-Thoss
Hamiltonian \cite{Meyer1979,Stock1997}, widely used in molecular
physics (see \cite{Dorta2012} and references therein). Notice
that, within this canonical formulation, the variables $I,\Phi$
play the role of generalized momentum and position, respectively.
Therefore, after a time re-scaling $t'\rightarrow 2 t$, the
solutions of $i\partial_{t}|\Psi\rangle=\hat H|\Psi\rangle$
($\hbar=1$) are the same as those of $\dot I=-\frac{\partial
H_{0}}{\partial \Phi}$ and $\dot \Phi=\frac{\partial
H_{0}}{\partial I}$ (the new time variable is again denoted as
$t$). Thus, the qubit can be taken as a classical particle moving
on the surface of $S^{2}$, as stated before. It is well known that
the motion of classical particles can be geometrized according to
the following theorems \cite{Arnoldbook}:

{\it Theorem 1.} A point mass confined to a smooth Riemannian
manifold moves along geodesic lines.

{\it Theorem 2}. In the case where there is a potential energy, it
can be shown that the trajectories of the equations of dynamics
are also geodesics in a certain Riemannian metric.

Therefore, one could ask wether similar theorems hold for the case
of a qubit. For example, if it is taken as a free classical
particle moving over $S^{2}$, then Hamilton's equations derived
from $H_{0}$ have to be the same as that of the geodesics of
$S^{2}$ written in action--angle coordinates. Although the qubit
trajectories coincide with the geodesics of $S^3$, in this  brief
article we show that this is not the case for $S^2$, as also
pointed out by Kryukov \cite{Kryukov2007}. However, these
trajectories are shown to be geodesics in a certain Riemannian
metric. In this sense, we extend the previous theorems to the
isolated qubit, which can be considered as a paradigmatic example
in quantum mechanics. Moreover, as the Euler
characteristic of the manifolds whose geodesics are the qubit
trajectories (qubit manifolds) is zero, it will be shown that they can
also be endowed with a Lorentzian metric, whose physical
interpretation is briefly discussed. In addition, these results
will also be extended, when possible, to non--isolated qubits by
means of an effective Hamiltonian description which includes
dissipative terms due to the presence of an environment
\cite{Bargueno2013}.

\section{Riemannian Qubit manifolds}

Let us start by defining a Riemannian qubit manifold.

{\it Definition 1.} Let $\mathcal{M}$ be a two--dimensional
connected, compact and orientable Riemannian $C^{n}$--manifold
($n\geq 2$) and let $H_{0}(u,v)$ be a Hamiltonian function for a
qubit, where $(u,v)$ are any pair of coordinates used to represent
$H_{0}$. If $\ddot u = -\frac{d}{dt}\left(\frac{\partial
H_{0}}{\partial v}\right)=f(u,v,\dot u,\dot v)$ and $\ddot v
=\frac{d}{dt}\left(\frac{\partial H_{0}}{\partial u}\right)
=g(u,v,\dot u,\dot v)$ coincide with the geodesics of
$\mathcal{M}$, then $\mathcal{M}$ is said to be a qubit manifold.

{\it Proposition 1.} No qubit manifold exists such that $(A_{i},A_{z})\ne (0,0)$ ($i=x$ or $y$) or
$A_{z}\ne 0$ and $(A_{x},A_{y})\ne (0,0)$.

{\it Proof}. The corresponding equations of motion issued from
$H_{0}$ are, in action--angle coordinates covering the region
$I\in (-1,1)$ and $\Phi\in[0,2 \pi]$,
\begin{eqnarray}
\label{eqs1} \dot I &=&2\sqrt{1-I^2}(A_{x}\sin \Phi-A_{y}\cos\Phi)\nonumber \\
\dot \Phi &=&-\frac{2I}{\sqrt{1-I^2}}(A_{y}\sin\Phi+A_{x}\cos\Phi)+ 2 A_{z}.
\end{eqnarray}
\noindent Therefore,
\begin{eqnarray}
\label{eqI}
&&\ddot I + \frac{I}{1-I^{2}}\dot I^{2} + \frac{1-I^{2}}{I}\dot \Phi \left(\dot \Phi-2A_{z}\right)=0 \nonumber \\
&& \ddot \Phi +\dot I \dot \Phi \frac{I^{2}+1}{I(I^{2}-1)}+\frac{2\dot I A_{z}}{I(1-I^{2})}=0.
\end{eqnarray}
\noindent Thus, if the latter pair of equations is likely to
describe the geodesics of $\mathcal{M}$ then, by comparing them
with the geodesic equation
\begin{equation}
\label{eqgeo}
\ddot x^{\mu}+\Gamma^{\mu}_{\nu \delta}\dot x^{\nu}\dot x^{\delta}=0,
\end{equation}
\noindent it has to be $A_{z}=0$. Thus, no qubit manifold exist
such that $(A_{i},A_{z})\ne (0,0)$ ($i=x$ or $y$) or
$A_{z}\ne 0$ and $(A_{x},A_{y})\ne (0,0)$
in action--angle
coordinates. Moreover, by defining a new pair of coordinates,
$u=f(I,\Phi)$ and $v=g(I,\Phi)$,  a new $A_{z}$--term linear in
$\dot u$ and $\dot v$ appears. Therefore, no qubit manifold exist
such that $(A_{i},A_{z})\ne (0,0)$ ($i=x$ or $y$) or
$A_{z}\ne 0$ and $(A_{x},A_{y})\ne (0,0)$. $\square$

In the following, the qubit manifold corresponding to the case
$A_{i}\ne 0$ ($i=x$ or $y$) and $A_{z}=0$ will be denoted by
$\mathcal{M}_{x}$ or $\mathcal{M}_{y}$. If $A_{x}\ne 0$, $A_{y}\ne 0$ and $A_{z}=0$, it will be denoted by $\mathcal{M}_{xy}$. Finally,
in the $(A_{x},A_{y})=(0,0)$ and $A_{z}\ne 0$ case, it will be denoted by
$\mathcal{M}_{z}$.

Although an easy but lengthy calculation shows that no metric
connection exist on $\mathcal{M}_{x}$, $\mathcal{M}_{y}$ nor $\mathcal{M}_{xy}$ such
that it is diagonal neither in action--angle $(I,\Phi)$ nor spherical
($\theta,\Phi$) coordinates $(I=\nobreak \cos \theta)$, the
following propositions can be stated.

{\it Proposition 2.} The metric
\begin{equation}
ds^{2}=-\frac{1}{2 I}\frac{1+I^{2}}{1-I^{2}}dI^{2}
+2 \sqrt{\frac{1+I^{2}}{I(1-I^{2})(I^{2}-1)}}d\Phi dI 
+\frac{\sqrt{I^{2}-1}}{I}d\Phi^{2}
\end{equation}
\noindent is a metric for $\mathcal{M}_{x}$, $\mathcal{M}_{y}$ and $\mathcal{M}_{xy}$
in action--angle coordinates.

{\it Proof}. If $A_{z}=0$, by comparing Eqs. (\ref{eqI}) with
Eqs. (\ref{eqgeo}), the only nonvanishing connection
coefficients are shown to be given by
\begin{eqnarray}
\label{Christ}
\Gamma^{I}_{II}&=&\frac{I}{1-I^{2}} \nonumber \\
\Gamma^{I}_{\Phi \Phi}&=&\left(\Gamma^{I}_{II}\right)^{-1} \nonumber \\
\Gamma^{\Phi}_{I\Phi}&=&\frac{1}{2I} \frac{I^{2}+1}{I^{2}-1}
\end{eqnarray}
\noindent Let us impose that these are the Levi--Civita
connections coefficients for a metric of the general form
$ds^{2}=\nobreak f(I)dI^{2}+2g(I)dI d\Phi+h(I)d\Phi^{2}$. Then,
using the well--known relation between the Christoffel symbols and the metric coefficients,
$\Gamma^{\lambda}_{\mu\nu}=\frac{1}{2}g^{\lambda\rho}
\left(g_{\rho\mu,\nu}+g_{\rho\nu,\mu}-g_{\mu\nu,\rho} \right)$, we
arrive at
\begin{eqnarray}
\label{eqsth2}
&&\frac{1}{2f}\frac{df}{dI}+\frac{1}{g}\frac{dg}{dI}= \frac{I}{1-I^{2}}\nonumber \\
&& -\frac{1}{2f}\frac{dh}{dI}=\frac{1-I^{2}}{I}\nonumber \\
&& \frac{1}{2h}\frac{dh}{dI}=\frac{1}{2I} \frac{I^{2}+1}{I^{2}-1}.
\end{eqnarray}
\noindent whose solutions are given by
\begin{eqnarray}
f(I)&=&-\frac{1}{2 I}\frac{1+I^{2}}{1-I^{2}}\nonumber \\
g(I)&=&\sqrt{\frac{1+I^{2}}{I(1-I^{2})(I^{2}-1)}}\nonumber \\
h(I)&=&\frac{\sqrt{I^{2}-1}}{I}.
\end{eqnarray}
\noindent Then, the desired result is proved. $\square$

{\it Proposition 3}. There exists a pair of coordinates $(\bar I,\bar \Phi)$ such that
\begin{equation}
ds^{2}=2\,d\bar I d\bar \Phi + h(\bar I)d\bar \Phi^{2}
\end{equation} 
is a metric for $\mathcal{M}_{x}$, $\mathcal{M}_{y}$ and $\mathcal{M}_{xy}$.

{\it Proof}. This result follows from the definition of the new pair of coordinates $(\bar I,\bar \Phi)$, given by
\begin{eqnarray}
\bar \Phi &=& \Phi + \int \frac{g(I)-\sqrt{-{\mathrm{det}}}}{h(I)}dI \nonumber \\
\bar I &=& \int \sqrt{-{\mathrm{det}}}\, dI,
\end{eqnarray}
where $\mathrm{det}=f(I)h(I)-g^{2}(I)$. $\square$

Now it is interesting and illustrative to analyze the case for
only $A_z \neq 0$. Thus,

{\it Proposition 4}. The qubit manifold $\mathcal{M}_{z}$ is the compact flat cylinder $[-1,1]\times S^{1}$.

{\it Proof}. In this case, the dynamics is given by
\begin{eqnarray}
\label{eqs3} \dot I &=&0\nonumber \\
\dot \Phi &=&2 A_{z}
\end{eqnarray}
\noindent or, in terms of spherical coordinates,
\begin{eqnarray}
\label{eqs4} &&\ddot \theta +\dot \theta^{2} \cot \theta=0\nonumber \\
&&\ddot \Phi =0.
\end{eqnarray}
If the latter pair of equations describe the geodesics for some
metric then, by comparing again them with Eqs. (\ref{eqgeo}), the
non-vanishing connection coefficients are given by
\begin{equation}
\Gamma^{\theta}_{\theta\theta}= \cot \theta \;\;\; \mathrm{and}\;\;\; \Gamma^{\Phi}_{ij}= 0.
\end{equation}
Using the fact that any two dimensional Riemannian metric can be
locally recast as $ds^{2}=\nobreak du^{2}+f(u,v)dv^{2}$, let us
look for a metric of the general form $ds^{2}=\nobreak
f(\theta,\Phi)d\theta^{2}+d\Phi^{2}$. If the metricity condition
of the connection is imposed, the differential equation which has
to be fulfilled is $\frac{1}{2f}\frac{\partial f}{\partial
\theta}=\cot \theta$. After simple manipulations, its solution is
given by $f(\theta)=\sin^{2}\theta$. Then, the corresponding
metric is
\begin{equation}
ds^{2}=\sin^{2}\theta d\theta^{2} + d\Phi^{2}
\end{equation}
\noindent or, in action--angle coordinates,
\begin{equation}
ds^{2}=dI^{2}+ d\Phi^{2}
\end{equation}
\noindent which describes a flat cylinder embedded in $\mathbb{R}^{3}$.
$\square$

\begin{figure}[h!]
\begin{center}
\includegraphics*[height=10cm,width=7cm]{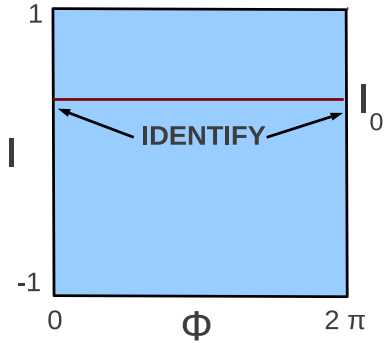}
\end{center}
\vspace{-3.0cm} \caption{\label{fig1}Geodesic of
$\mathcal{M}_{z}$ (red line) in the
$(I,\Phi)$--plane with constant $I_{0}$.}
\end{figure}

Notice that this result is consistent with the fact that the
solution of Eqs. (\ref{eqs3}) is a straight line in the
$(I,\Phi)$--plane which, after identifying $\Phi(0)$ with
$\Phi(2\pi)$ becomes a flat cylinder (see Fig. (\ref{fig1})).

{\it Proposition 5}. The qubit manifolds $\mathcal{M}_{x}$, $\mathcal{M}_{y}$ and $\mathcal{M}_{xy}$ are the compact and curved 
cylinders with $[-1,1]\times S^{1}$ topology.

{\it Proof}. On one hand, it can be shown by Proposition 3 that the metric for $\mathcal{M}_{z}$, which is a compact and flat 
cylinder, can be recast as $ds^{2}=\nobreak 2d\bar I d\bar \Phi + d\bar \Phi^{2}$. On the other hand, a straightforward calculation shows that 
the scalar curvature corresponding to the metric of the form
$ds^{2}= \nobreak 2\,d\bar I d\bar \Phi + h(\bar I)d\bar \Phi^{2}$ is given by $R=h''(\bar I)$, where a prime denotes $d/d\bar I$. Therefore,
the qubit manifolds $\mathcal{M}_{x}$, $\mathcal{M}_{y}$ and $\mathcal{M}_{xy}$ can be taken to be curved cylinders. $\square$

We note that any qubit manifold is conformally flat (in fact, any two--dimensional
Riemannian manifold is conformally flat \cite{Nakaharabook}). Although this is not evident for the
$\mathcal{M}_{x}$, $\mathcal{M}_{y}$ nor $\mathcal{M}_{xy}$ cases, this can be proved by inspection for the $\mathcal{M}_{z}$ qubit 
manifold since it corresponds to a flat cylinder.




\section{Extension to open--system dynamics}

In a previous work, a geometrical description of a
Caldeira--Legget--like open system dynamics for a qubit has been
developed \cite{Bargueno2013}, showing that the effective
open--system dynamics driven by the Hamiltonian
\begin{equation}
H = H_{0}+
\frac{1}{2}\sum_{i}\left(p^{2}_{i}+x^{2}_{i}\omega_{i}^{2}\right)
-\Phi\sum_{i}c_{i}x_{i}+\sum_{i}\Phi^{2}c^{2}_{i} \label{hcal}
\end{equation}
where the oscillator mass has been taken to be one and $c_{i}$ are
the system--bath coupling constants, can be described by
\cite{Dorta2012,Pedro2013}
\begin{equation}
H_{t}=H_{0}+2\gamma\Phi
\dot \Phi -\xi \Phi,
\end{equation}
where $\gamma$ stands for a friction constant and $\xi$ is a
stochastic Gaussian process representing a noisy environment (for
technical details see \cite{Bargueno2013} and references therein).
We remark that the effective Hamiltonian function, $H_{t}$, is not
a conserved quantity (notice the $t$ index in $H_{t}$). After a
second time derivative, the corresponding equations of motion
issued from $H_{t}$ can be written as
\begin{eqnarray}
\label{eqII}
\ddot I &+& \frac{I}{1-I^{2}}\dot I^{2} + \frac{1-I^{2}}{I}
\dot \Phi \left(\dot \Phi-2A_{z}\right)
\nonumber \\
&-&2\gamma \left(\frac{I^{2}+1}{I(I^{2}-1)}\dot I \dot \phi +
\frac{2 A_{z} \dot I}{I}\right)+\dot {\xi}(t)=0 \nonumber \\
&& \ddot \Phi +\dot I \dot \Phi \frac{I^{2}+1}{I(I^{2}-1)}+\frac{2\dot I A_{z}}{I(1-I^{2})}=0.
\end{eqnarray}

These equations (or the existence of $H_{t}$) motivate the
following definition:

{\it Definition 2.} Let $\mathcal{M}^{\gamma}$ be a
two--dimensional connected, compact and orientable Riemannian
$C^{n}$--manifold ($n\geq 2$) and let $H_{t}\equiv H_{t}(u,v,\dot
u, \dot v)$ be an effective Hamiltonian function for a dissipative
qubit. The pair $(u,v)$ refers to any pair of coordinates used to
represent $H_{0}$. If $\ddot u = -\frac{d}{dt}\left(\frac{\partial
H_{t}}{\partial v}\right)=f(u,v,\dot u,\dot v)$ and $\ddot v
=\frac{d}{dt}\left(\frac{\partial H_{t}}{\partial u}\right)
=g(u,v,\dot u,\dot v)$ coincide with the geodesics of
$\mathcal{M}^{\gamma}$, then $\mathcal{M}^{\gamma}$ is said to be
a dissipative qubit manifold.

As carried out in the non--dissipative description, the
dissipative qubit manifold corresponding to the case $A_{i}\ne 0$
($i=x$ or $y$) and $A_{z}=0$ will be denoted by
$\mathcal{M}_{i}^{\gamma}$. In the $(A_{x},A_{y})\ne (0,0)$ and $A_{z}=0$ case, it will
be denoted by $\mathcal{M}_{xy}^{\gamma}$. Finally, in the $(A_{x},A_{y})=(0,0)$ and
$A_{z}\ne 0$ case, it will be denoted by
$\mathcal{M}_{z}^{\gamma}$. The only way this dynamics could
correspond to a geodesic motion is when noisy terms are not
included.

{\it Proposition 6.} No dissipative qubit manifold exists such
that $(A_{i},A_{z})\ne (0,0)$ ($i=x$ or $y$) and
$A_{z}\ne 0$ and $(A_{x},A_{y})\ne (0,0)$.

{\it Proof}. Similar to Proposition 1. Compare Eqs. (\ref{eqII})
(without the term of the time derivative of the noise) with the
geodesic equation. $\square$

{\it Proposition 7.} No dissipative qubit manifold exists such
that $A_{i} \ne 0$ ($i=x$ or $y$) and $A_{z}=0$.

{\it Proof}. If Eqs. (\ref{eqII}) (without the noisy term) are
likely to describe the geodesics of $\mathcal{M}_{x}^{\gamma}$, $\mathcal{M}_{y}^{\gamma}$
or $\mathcal{M}_{xy}^{\gamma}$, then the corresponding connection
coefficients are given by Eqs. (\ref{Christ}) together with
$\Gamma^{I}_{I\Phi}=2\gamma \Gamma^{\Phi}_{I\Phi}$. Thus, the
differential equations one has to solve are Eqs. (\ref{eqsth2})
together with $\frac{1}{f}\frac{dg}{dI}+\frac{1}{2g}\frac{dh}{dI}=-\gamma \frac{I^{2}+1}{I(I^{2}-1)}$.
The incompatibility of these equations proves the required result.
$\square$

In spite of these negative results, we have the following

{\it Proposition 8.} The dissipative qubit manifold
$\mathcal{M}_{z}^{\gamma}$ is the compact flat cylinder
with $[-1,1]\times S^{1}$ topology.

{\it Proof}. In this case, the corresponding dissipative dynamics
is given by
\begin{eqnarray}
\label{eqs3bis} \dot I &=&-\gamma \dot \Phi\nonumber \\
\dot \Phi &=&2 A_{z}
\end{eqnarray}
or, in terms of spherical coordinates,
\begin{eqnarray}
\label{eqs4bis} &&\ddot \theta +\dot \theta^{2} \cot \theta=0\nonumber \\
&&\ddot \Phi =0
\end{eqnarray}
which coincide with Eqs. (\ref{eqs4}). Therefore, the required result follows from Proposition
3. $\square$

\begin{figure}[h!]
\begin{center}
\includegraphics*[height=10cm,width=7cm]{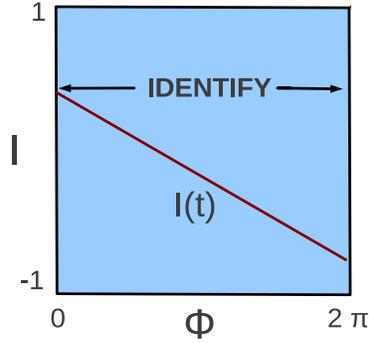}
\end{center}
\vspace{-3.0cm} \caption{\label{fig2}Geodesic of
$\mathcal{M}_{z}^{\gamma}$ (red line) in the
$(I,\Phi)$--plane with $I(t)=-\gamma \Phi(t)+I(0)+\gamma
\Phi(0)$.}
\end{figure}

The main difference with the non--dissipative case is that, in the
$\gamma\ne 0$ situation, the geodesic does not lie in the same
plane for all $t$. This can be shown by noting that the solutions
of Eqs. (\ref{eqs3bis}) give place to $I(t)=-\gamma
\Phi(t)+I(0)+\gamma \Phi(0)$, which becomes an helix after
identifying $\Phi(0)$ with $\Phi(2\pi)$ (see Fig. (\ref{fig2})).
\vspace{-2.0cm}

\vspace{3.0cm}
\section{Lorentzian qubit manifolds}

Extending some of the previous results to the Lorentzian case, far
from being a purely mathematical generalization, can be physically
justified. In particular, introducing a Lorentzian signature in
$\mathcal{M}_{z}$ seems to be rather natural since, in this case,
$H_{0}=2A_{z} I$ and $I$ is a generalized momentum, $p_{I}$. Then,
by taking $2A_{z}=c$, the Hamiltonian function can be recast as
$H_{0}=p_{I}c$, which is precisely the dispersion  relation of a
massless particle. Although there are global obstructions for a
manifold to admit a Lorentzian metric, this photon--like particle
will be shown to propagate in two--dimensional Minkowski space
with the cylinder topology. As in the Riemannian case, dissipation
will be included by adding the term $2\gamma\Phi\dot \Phi$ to
$H_{0}$.

The obstructions for a manifold to admit a Lorentzian metric are
reflected in the following theorem \cite{Ehrlichbook}:

{\it Theorem 3}. A manifold admits a Lorentzian metric if and only
if it is noncompact or has zero Euler characteristic.

Therefore, remembering that the only dissipative qubit manifold is
$\mathcal{M}_{z}^{\gamma}$, the following results can be stated:

{\it Proposition 9}. Any non--dissipative or dissipative qubit
manifold admits a Lorentzian metric.

{\it Proof}. As the cylinder has zero Euler characteristic then, by Propositions 4, 5 and Theorem 3, this result is
straightforward. $\square$

{\it Proposition 10.} $\mathcal{M}_{z}$ and
$\mathcal{M}_{z}^{\gamma}$ are qubit Lorentzian manifolds with the
compact flat cylinder $[-1,1]\times S^{1}$ topology.

{\it Proof}. Any two--dimensional Lorentzian metric can be locally
recast as $ds^{2}=du^{2}-f(u,v)dv^{2}$. Then, by adapting the
procedure employed in Propositions 4 and 8 to the Lorentzian case,
we reach that
\begin{equation}
ds^{2}=d\Phi^{2}-dI^{2}
\end{equation}
\noindent is a Lorentzian metric for $\mathcal{M}_{z}$ and
$\mathcal{M}_{z}^{\gamma}$. $\square$


{
%
%


\section{Conclusions and future work}

In this work, we have applied a geometrization of quantum
mechanics using the first Hopf fibration to show that the
trajectories of a qubit dynamics over the two--sphere are
geodesics in certain Riemannian or physically--sound Lorentzian
metrics, which turned to be flat and curved cylinders.
In addition, by including dissipative terms to the
dynamics by means of a Caldeira--Legget--like coupling to the
environment, the previous findings for the simplest dissipative
qubit have been generalized. Extension of these results to deal
with two--qubit entanglement on $S^{4}$ and, in general, with the
dynamics of n--level systems on $\mathbb{C}P^{n-1}$ is currently in
progress.

\section{Acknowledgements}

This work has been funded by the MICINN (Spain) through Grants No.
CTQ2008--02578 and FIS2011--29596-C02-01. P. B. acknowledges a
Juan de la Cierva fellowship from the MICINN, a J\'ovenes
Profesores e Investigadores fellowship from Banco Santander
(Spain) and H.-C. P-R and G. R--L acknowledge a scientific project
from INSTEC (Cuba). P. B. would like to express his gratitude  to
all members of the Instituto Superior de Tecnolog\'{\i}as y
Ciencias Aplicadas (La Habana, Cuba), where part of this work has
been done, for their kind hospitality.


\begin{thebibliography}{n}
\bibitem{Rosner2001}J. L. Rosner and S. A. Slezak, Am. J. Phys. {\bf 69}, 44 (2001).
\bibitem{Harris1978} R. A. Harris and L. Stodolsky, Phys. Lett. B {\bf 78}, 313 (1978).
\bibitem{Quack1986} M. Quack, Chem. Phys. Lett. {\bf 132}, 147 (1986).
\bibitem{Feynmanbook}R. P. Feynman, R. B. Leighton and M. Sands, Lectures on Physics vol. III,
 Addison--Wesley, Reading, MA (1965).
\bibitem{Ladd2010}T. D. Ladd, F. Jelezko, R. Laflamme, Y. Nakamura, C. Monroe and J. L. OBrien, Nature {\bf 464}, 45 (2010).
\bibitem{Nielsenbook}M. Nielsen, and I.L. Chuang, Quantum computation and quantum
information, Cambridge University Press, Cambridge (2011).
\bibitem{Kibble1979}T. W. B. Kibble, Comm. Math. Phys. {\bf 65} (2), 189 (1979).
\bibitem{Heslot1985}A. Heslot, Phys, Rev. D {\bf 31}, 1341 (1985).
\bibitem{Anandan1990}J. Anandan and Y. Aharonov, Phys. Rev. Lett. {\bf 65}, 1697 (1990).
\bibitem{Gibbons1992} G. W. Gibbons, Jour. Geom. Phys. {\bf 8}, 147 (1992).
\bibitem{Ashtekar1998}A. Ashtekar and T. A. Schilling, Geometrical formulation of quantum mechanics,
in A. Harvey (ed.), On Einstein's Path, Springer (1998).
\bibitem{Brody2001}D. C. Brody and L. P. Hughston, Jour. Geom. Phys. {\bf 38}, 19 (2001)
\bibitem{Chruszinskybook}D. Chruszinsky and A. Jamiolkovsky, Geometric Phases in Classical and Quantum Mechanics,
Progress in Mathematical Physics {\bf 36}, Birkh\"auser, Boston
(2004).
\bibitem{Hopf1931}H. Hopf, Matematische Annalen {\bf 104}, 637 (1931).
\bibitem{Urbantke1991}H. Urbantke, A. J. Phys. {\bf 59}, 503 (1991).
\bibitem{Mosseri2001}R. Mosseri and R. Dandoloff, J. Phys. A: Math. Gen. {\bf 34}, 10243 (2001).
\bibitem{Urbantke2003}H. K. Urbantke, J. Geom. Phys. {\bf 46}, 125 (2003).
\bibitem{Strocchi1966}F. Strocchi, Rev. Mod. Phys. {\bf 38}, 36 (1966).
\bibitem{Meyer1979}H.-D. Meyer and W. H. Miller, J. Chem. Phys. {\bf 70}, 3214 (1979).
\bibitem{Stock1997}G. Stock and M. Thoss, Phys. Rev. Lett. {\bf 78}, 578 (1997).
\bibitem{Dorta2012}A. Dorta--Urra {\it et al.}, J. Chem. Phys. {\bf 136}, 174505 (2012).
\bibitem{Arnoldbook} V. I. Arnold, Mathematical Methods of Classical Mechanics (2nd ed.),
Springer--Verlag (1989).
\bibitem{Kryukov2007} A. A. Kryukov, Found. Phys. {\bf 37}, 3
(2007).
\bibitem{Bargueno2013}P. Bargue\~no and S.Miret--Art\'es, Phys. Rev. A {\bf 87}, 012125 (2013).
\bibitem{Nakaharabook} M. Nakahara, Geometry, Topology and Physics, Institute of Physics Publishing, Bristol (2003).
\bibitem{Pedro2013}H. C. Pe\~nate-Rodr\'{\i}guez {\it et al.}, Chirality {\bf 25}, in press (2013).
\bibitem{Ehrlichbook} J. K. Beem, P. E. Ehrlich and K. L. Easley, Global Lorentzian Geometry,
Pure and Applied Mathematics {\bf 202}, Marcel Dekker (1996).

\end{thebibliography}
\end{document}